\documentclass[12pt, a4paper]{article}
\usepackage{graphicx}
\begin{document}
\title{\bf\boldmath Experimental Study of $\rho\to\pi^0\pi^0\gamma$ and
$\omega\to\pi^0\pi^0\gamma$ Decays}
\author{
M.N.Achasov, K.I.Beloborodov, A.V.Berdyugin, A.G.Bogdanchikov,\\
A.V.Bozhenok, D.A.Bukin, S.V.Burdin, A.V.Vasiljev, V.B.Golubev, \\ 
T.V.Dimova, V.P.Druzhinin\thanks{e-mail: druzhinin@inp.nsk.su},
V.N.Ivanchenko, A.A.Korol, I.A.Koop, \\
S.V.Koshuba, A.V.Otboev, E.V.Pakhtusova, A.A.Salnikov,\\
S.I.Serednyakov, V.A.Sidorov, Z.K.Silagadze, A.N.Skrinsky,\\
A.G.Skripkin, Yu.V.Usov, V.V.Shary, Yu.M.Shatunov.
\vspace{1mm}\\ Budker Institute of Nuclear Physics, \\ 
Novosibirsk State University,\\
Novosibirsk 630090, Russia} 
\date{}
\maketitle
\begin{abstract}
The $e^+e^-\to\pi^0\pi^0\gamma$ process was studied in the SND
experiment at VEPP-2M $e^+e^-$ collider in the energy region
0.60--0.97~GeV. From the analysis of the energy dependence of
measured cross section the branching ratios
$B(\omega\to\pi^0\pi^0\gamma)=  (6.6^{+1.4}_{-0.8}\pm0.6)\cdot10^{-5}$
and $B(\rho\to\pi^0\pi^0\gamma)=(4.1^{+1.0}_{-0.9}\pm0.3)\cdot10^{-5}$
were obtained.
\vspace{1mm}\\
{\bf PACS: 13.65.+i,14.40.Cs}
\end{abstract}
\section{Introduction}
In 1998 and 2000 the experiments with Spherical Neutral Detector
(SND)\cite{SND} at VEPP-2M $e^+e^-$ collider were carried out
in the energy range $E=360-970$~MeV where cross section of
$e^+e^-$ annihilation into hadrons
is determined by the $\rho$ and $\omega$ meson decays.
The integrated luminosity of 9 pb$^{-1}$ collected in the experiment
corresponds to $3.5\cdot10^6$ and $7\cdot10^6$
produced $\rho$ and $\omega$ mesons, respectively.
One of the goals of the experiment was the investigation of the rare process
\begin{equation}
e^+e^-\to\rho,\omega\to\pi^0\pi^0\gamma.
\label{ppg}
\end{equation}
Our preliminary study~\cite{rhoppg} of the process (\ref{ppg}) was based
on 1/3 of collected statistics. Its results were 
the first measurement of
$B(\rho\to\pi^0\pi^0\gamma)=(4.8^{+3.4}_{-1.8}\pm0.2)\cdot10^{-5}$
and the measurement of
$B(\omega\to\pi^0\pi^0\gamma)=(7.8\pm2.7\pm2.0)\cdot10^{-5}$
confirming the only previous measurement of this decay by GAMS:
$B(\omega\to\pi^0\pi^0\gamma)=(7.2\pm2.5)\cdot10^{-5}$~\cite{GAMS}.

\begin{figure}[t]
\centering
\includegraphics[width=0.98\textwidth]{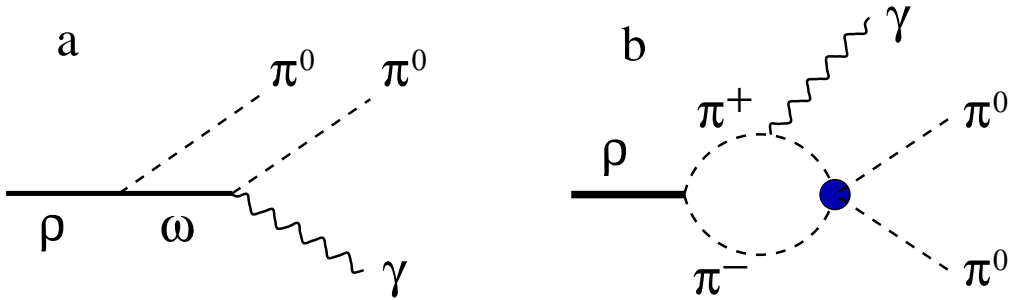}
\caption{\small
Two mechanism of $\rho\to\pi^0\pi^0\gamma$ decay: (a) is the VMD mechanism,
(b) is the transition through pion loop.} 
\label{diad}
\end{figure}

The theoretical study of the $\rho,\omega\to\pi^0\pi^0\gamma$ decays
was begun by P.Singer in Ref.~\cite{Singer1} where the transitions via
$\omega\pi^0$ (Fig.~\ref{diad}a) and $\rho^0\pi^0$ intermediate states were
suggested. The vector meson dominance (VMD)
calculation with these intermediate states leads to branching ratios
$\sim1 \cdot10^{-5}$ and $\sim 3\cdot10^{-5}$ for
$\rho\to\pi^0\pi^0\gamma$ and $\omega\to\pi^0\pi^0\gamma$,
respectively~\cite{Bramon1}.
\begin{table}[b]
\centering
\caption{\small The branching ratios for different contributions to
$\rho\to\pi^0\pi^0\gamma$ decay. The chiral loop or $\rho\to\sigma\gamma$ 
contribution is calculated in the framework of 
Chiral Perturbation Theory ($\chi PT$)~\cite{Bramon3,Bramon2},
Unitarized Chiral Perturbation Theory ($U\chi PT$)~\cite{Marco,Oset}, 
Linear Sigma Model ($L \sigma M$)~\cite{Bramon2} and 
$\sigma$ pole model~\cite{Gokalp1}.\vspace{2mm}} 
\label{theort}
\begin{tabular}{|l|c|c|c|c|c|c|c|}
\hline
model & VMD & loops, $\sigma\gamma$ & total \\
\hline
$\chi PT$ & $1.3\cdot10^{-5}$ & $1.0\cdot10^{-5}$ & $2.9\cdot10^{-5}$ \\ 
$U\chi PT$ & $1.5\cdot10^{-5}$ & $1.5\cdot10^{-5}$ & $4.2\cdot10^{-5}$ \\ 
$L \sigma M$ & $1.3\cdot10^{-5}$ & $(0.8-2.1)\cdot10^{-5}$ & $(2.8-4.7)\cdot10^{-5}$ \\ 
$\sigma$ pole & $1.0\cdot10^{-5}$ & $\sim2\cdot10^{-3}$ & $\sim2\cdot10^{-3}$ \\ 
\hline
\end{tabular}
\end{table}

For $\rho\to\pi^0\pi^0\gamma$ decay another mechanism
through the pions loops (Fig.~\ref{diad}b) is also possible~\cite{Bramon1}.
The branching ratios expected for this mechanism in different
models~\cite{Bramon3,Bramon2,Marco,Oset} are listed in Table~\ref{theort}.
It was noted in Ref.~\cite{Marco} that the $\rho\to\pi^0\pi^0\gamma$ decay
via chiral loops
can be interpreted as $\rho\to\sigma\gamma$, where $\sigma$ is
a scalar state decaying into $\pi\pi$ pair.
The dependence of $B(\rho\to\pi^0\pi^0\gamma)$ on $\sigma$ parameters 
was studied in Ref.~\cite{Bramon2}.
The range of $B(\rho\to\pi^0\pi^0\gamma)$ values in L$\sigma M$ model
corresponds to different $\sigma$ widths.
Since the amplitudes of
$\rho\to\omega\pi^0\to\pi^0\pi^0\gamma$ transition (Fig.~\ref{diad}a) and 
of the pion loops (Fig.~\ref{diad}b)
are of the same order of magnitude, their interference is substantial. The
interference  contribution into branching ratio is predicted to be positive. 
The theoretical values for the total branching ratios are also listed in 
Table~\ref{theort}. The predictions of the chiral 
models~\cite{Bramon3,Bramon2,Marco,Oset}
are in agreement with our previous experimental result.
The significantly larger value of $B(\rho\to\pi^0\pi^0\gamma)$ was
obtained in Ref.~\cite{Gokalp1} using $\sigma$ pole model. Their
result contradicts to existing experimental data.

In $\omega\to\pi^0\pi^0\gamma$ decay the contribution of
pion loops is \textsl{G}-parity suppressed while the contribution of
kaon loop is small due to large kaon mass.
Therefore, it is assumed that the $\omega\to\pi^0\pi^0\gamma$ decay 
proceeds through $\rho^0\pi^0$ intermediate state. 
The first measurement
$B(\omega\to\pi^0\pi^0\gamma)=(7.2\pm2.5)\cdot10^{-5}$~\cite{GAMS}
significantly exceeded the existent prediction of VMD model:
$3\cdot10^{-5}$~\cite{Bramon1}.
An attempt to explain this discrepancy was done in Ref.~\cite{Singer2}
where $\rho\,$--$\,\omega$ mixing
was taken into account and coupling constants were extracted from
experimental values of
$\Gamma(\omega\to 3\pi)$ and $\Gamma(\rho^0\to \pi^0\gamma)$.
As a result the
estimated value of $B(\omega\to\pi^0\pi^0\gamma)$ increased up to
$(4.6\pm1.1)\cdot 10^{-5}$. Similar results (4.5--4.7)$\cdot 10^{-5}$
were then obtained in Refs.~\cite{Bramon3, Oset}. In Ref.~\cite{Gokalp2}
the large experimental value of $B(\omega\to\pi^0\pi^0\gamma)$ was
explained by  additional contribution of the $\omega\to\sigma\gamma$
transition and used to extract the value of $g_{\omega\sigma\gamma}$
coupling constant. 

In the present work we present the experimental results on
$B(\omega\to\pi^0\pi^0\gamma)$ and $B(\rho\to\pi^0\pi^0\gamma)$
based on full SND data sample. 
\section{Event selection}
 For analysis five-photon events with the energy deposition in the 
calorimeter
\begin{equation}
E_{tot}>0.7\cdot E
\label{etot}
\end{equation}
and the total momentum measured by the calorimeter 
\begin{equation}
P_{tot}<0.15\cdot E/c
\label{ptot}
\end{equation}
were selected. Here $E$ is $e^+e^-$ center of mass energy.

Due to high beam background rate
in 5\% of events fake photons appear. This makes possible
for lower photon multiplicity QED processes
$e^{+}e^{-} \to 2\gamma,\,3\gamma,4\gamma$, and
$\rho,\omega\to \pi^0\gamma, \eta\gamma \to 3 \gamma$ decays to imitate
five-photon events producing
main background contribution for the process under study.
Detector response to the beam background was studied using 
special events recorded with a random generator trigger. The information
on the fired detector channels in these events was used for simulation of
the process under study and the background processes.
Considerable suppression (by a factor of 8) of the background from
events with fake photons was achieved by imposing the following cuts:
\begin{equation}
E_{min} > 30 \,\mbox{MeV},\:30^{\circ}<\theta_{min}<150^{\circ},
\label{emin}
\end{equation}
where $E_{min}$ and $\theta_{min}$ are the energy and polar angle of the 
softest photon in an event.
These cuts reduce the detection efficiency for the process under study by
25\%.
Another background source is the
$e^{+}e^{-} \to\eta\gamma \to 3\pi^0\gamma \to 7\gamma$ reaction producing
five-photon  events mainly due to the merging of near
photons. To suppress this background, the parameter $\chi_{\gamma}$
describing transverse energy deposition profile
of the detected photon~\cite{xinm} was used. The cut
\begin{equation}
\chi_{\gamma}<5
\label{xinm}
\end{equation}
suppresses the $e^{+}e^{-}\to\eta\gamma$ background
by a factor of 2 with a 5\% loss of actual 5-photon events.

\begin{figure}[t]
\begin{minipage}[t]{0.47\textwidth}
\includegraphics[width=0.98\textwidth]{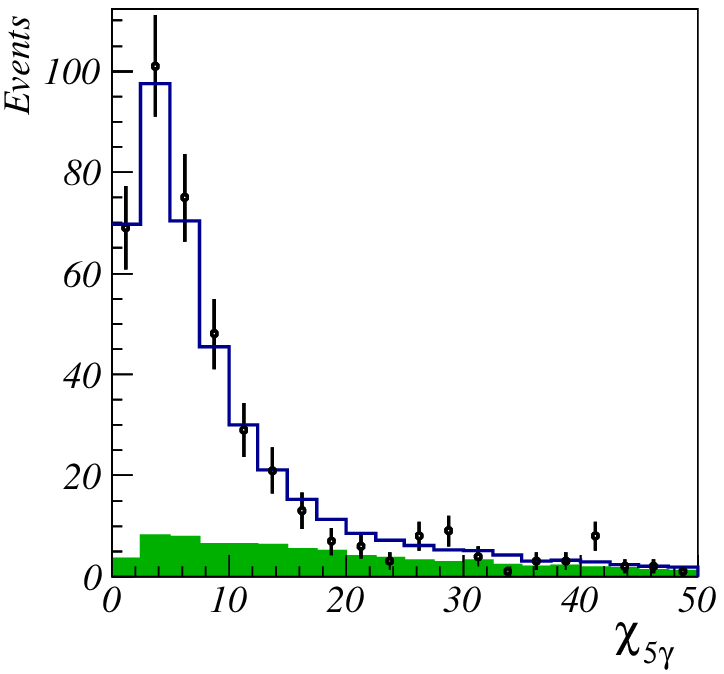}
\caption{\small The $\chi_{5\gamma}$ distribution.
Points with error bars represent experimental data. The 
histogram is a simulation of the process (\ref{ppg}) and
background processes. The shaded histogram shows the contribution of
background processes.
}
\label{f1}
\end{minipage}
\hfill
\begin{minipage}[t]{0.47\textwidth}
\includegraphics[width=0.98\textwidth]{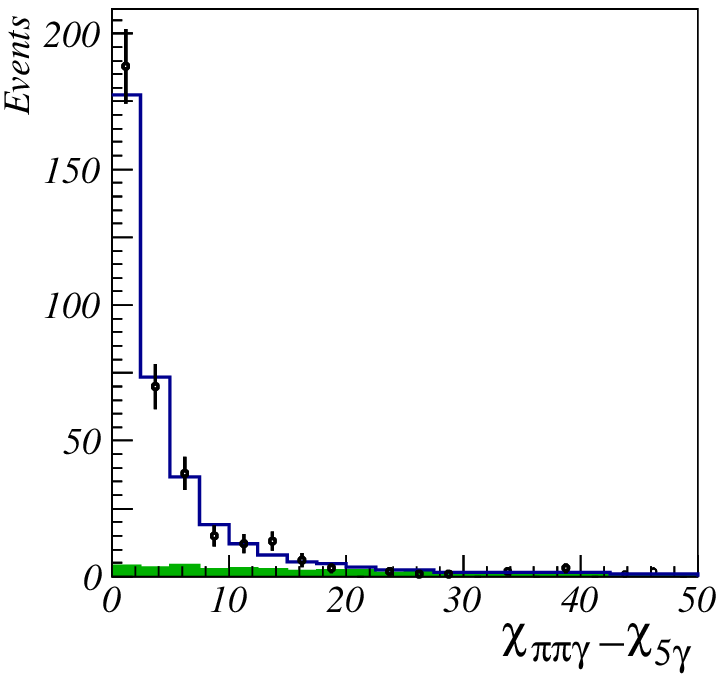}
\caption{\small The 
$\chi_{\pi\pi\gamma}-\chi_{5\gamma}$ distribution.
The points with error bars are experimental data. The 
histogram is the sum of simulated events of the process (\ref{ppg}) and
background processes. The shaded histogram is the contribution of
the background processes.
}
\label{f2}
\end{minipage}
\end{figure}

Further selection was based on the kinematic fitting of the events.
Compatibility of the event kinematics with
$e^{+}e^{-} \to 5\gamma$ and $e^{+}e^{-} \to 3\gamma$ hypotheses
was checked. For the $3\gamma$ hypothesis two out of five photons
were considered spurious: all 3-$\gamma$ subsets
were tested and the best one with minimum $\chi^2$ value was
selected. As a result of kinematic fitting the $\chi^2$ values,
$\chi_{5\gamma}$ and $\chi_{3\gamma}$, were calculated for
both hypotheses. The cut
\begin{equation}
\chi_{3\gamma}>20
\label{xi3g}
\end{equation}
practically eliminates $e^{+}e^{-} \to 2\gamma,\,3\gamma$ background
with the loss only 2.5\% of the events of the process under study.
Figure~\ref{f1} depicts the $\chi_{5\gamma}$ distribution of
the experimental and simulated events. The following cut
was imposed on this parameter:
\begin{equation}
\chi_{5\gamma}<20.
\label{xi5g}
\end{equation}
Finally, the events with two $\pi^0$ mesons were selected.
To do this the kinematic fit in $e^{+}e^{-} \to \pi^0\pi^0\gamma$
hypothesis was performed and the following cut was imposed:
\begin{equation}
\chi_{\pi\pi\gamma}-\chi_{5\gamma}<10.
\label{xifg}
\end{equation}
Here $\chi_{\pi\pi\gamma}$ is the $\chi^2$ value of the kinematic fit for 
the $e^{+}e^{-} \to \pi^0\pi^0\gamma$ hypothesis.
The $\chi_{\pi\pi\gamma}-\chi_{5\gamma}$
distributions for the experimental and simulated events
are shown in Fig.\ref{f2}.

The difference
between distributions of the background events and events of the process
under study (Figs. \ref{f1} and \ref{f2}) was used to
estimate the accuracy of the background calculation.
The experimental distribution  was fitted by the sum of
simulated distributions for the process (\ref{ppg}) and background.
As a result of the fit the ratio
$K=N_{exp}^{bkg}/N_{calc}^{bkg}=0.7\pm0.2$ was determined.
For softer selection criteria
without cuts (\ref{emin},\ref{xi3g},\ref{xinm}) this ratio is
equal to $1.30 \pm 0.05$ and does not depend on the beam
energy at our level of statistical accuracy.
From this we conclude that the systematic error of the background
estimation does not exceed 50\%.

The total of 310 events were selected with estimated background of
$15\pm7$ events. The main sources of the residual background are
$e^+e^-\to\eta\gamma$ and $e^+e^-\to 4\gamma$ processes.
The distribution of selected events and calculated background
over center of mass energy is given in the Table~\ref{t1}.
No events were detected below
600~MeV. The uncertainty of the center of mass 
energy, integrated luminosity, and detection efficiency are
listed in Table~\ref{t1} for each energy point. 
The uncertainty of the center of mass energy includes
the beam energy spread and the energy shift between 1998 and
2000 scans. The integrated luminosity was measured using
$e^+e^-\to\gamma\gamma$ process. The statistical error of the luminosity
in each energy point does not exceed 1\% and is not included in the table.
Its systematic error was estimated to be 3\%.
The detection efficiency for the process (\ref{ppg}) was determined by
simulation. The differential cross section of the
$e^+e^-\to\pi^0\pi^0\gamma$ process calculated in VMD model~\cite{ompn}
was used for simulation.
The systematic error of the detection efficiency, including the model
error due to
possible contribution from $\sigma\gamma$ intermediate state was estimated to
be 5\%.
\renewcommand{\arraystretch}{1.3}
\begin{table}[p]
\centering
\caption{\small
Center of mass energy $E$, its standard deviation $\sigma_E$,
integrated luminosity $L$,
number of selected events $N_{exp}$, calculated number of background
events $N_{bkg}$,
detection efficiency $\varepsilon$, radiative correction $1+\delta$
and the $e^+e^-\to\pi^0\pi^0\gamma$ cross section $\sigma^{exp}$.
\vspace{1mm}}
\label{t1}
\begin{tabular}{|l|c|r|r|c|c|c|c|}
\hline
 $E$,MeV &$\sigma_E$,MeV & L, nb$^{-1}$ & $N_{exp}$ & $N_{bkg}$ &
 $\varepsilon$ & $1+\delta$ & $\sigma^{exp}$,nb\\ 
\hline
 600.1 & 0.29 &  88.3 &   0 & 0.1 &0.273& 0.912 & $-0.005_{-0.002}^{+0.052}$\\
 630.1 & 0.30 & 116.1 &   0 & 0.1 &0.269& 0.906 & $-0.004_{-0.002}^{+0.041}$\\
 660.2 & 0.25 & 271.6 &   2 & 0.3 &0.273& 0.900 & $ 0.025_{-0.019}^{+0.040}$\\
 690.2 & 0.29 & 167.2 &   2 & 0.2 &0.263& 0.895 & $ 0.046_{-0.033}^{+0.067}$\\
 720.3 & 0.26 & 588.5 &   1 & 0.8 &0.251& 0.892 & $ 0.002_{-0.007}^{+0.018}$\\
 750.2 & 0.32 & 219.0 &   3 & 0.1 &0.259& 0.897 & $ 0.057_{-0.032}^{+0.057}$\\
 760.2 & 0.31 & 238.9 &   2 & 0.3 &0.251& 0.896 & $ 0.032_{-0.024}^{+0.049}$\\
 764.2 & 0.32 & 250.4 &   5 & 0.2 &0.254& 0.892 & $ 0.085_{-0.038}^{+0.060}$\\
 770.2 & 0.31 & 284.4 &   8 & 0.3 &0.253& 0.877 & $ 0.122_{-0.044}^{+0.063}$\\
 774.2 & 0.34 & 217.1 &   7 & 0.2 &0.252& 0.855 & $ 0.145_{-0.055}^{+0.081}$\\
 778.1 & 0.34 & 247.9 &   6 & 0.5 &0.261& 0.820 & $ 0.104_{-0.045}^{+0.068}$\\
 780.2 & 0.35 & 319.5 &  16 & 1.2 &0.263& 0.807 & $ 0.218_{-0.059}^{+0.075}$\\
 781.1 & 0.33 & 339.6 &  20 & 1.2 &0.267& 0.807 & $ 0.257_{-0.061}^{+0.076}$\\
 782.1 & 0.31 & 656.3 &  34 & 1.0 &0.257& 0.815 & $ 0.240_{-0.042}^{+0.050}$\\
 783.2 & 0.30 & 473.4 &  30 & 2.0 &0.253& 0.833 & $ 0.280_{-0.055}^{+0.066}$\\
 784.2 & 0.32 & 346.2 &  24 & 0.7 &0.261& 0.857 & $ 0.301_{-0.063}^{+0.077}$\\
 785.3 & 0.24 & 212.3 &  12 & 0.4 &0.257& 0.890 & $ 0.238_{-0.070}^{+0.094}$\\
 786.1 & 0.33 & 267.7 &  11 & 0.4 &0.255& 0.914 & $ 0.170_{-0.052}^{+0.071}$\\
 790.1 & 0.34 & 191.4 &   4 & 0.3 &0.258& 1.006 & $ 0.074_{-0.039}^{+0.064}$\\
 794.2 & 0.34 & 206.7 &   1 & 0.2 &0.256& 1.044 & $ 0.014_{-0.015}^{+0.042}$\\
 800.2 & 0.32 & 276.8 &  10 & 0.3 &0.255& 1.053 & $ 0.130_{-0.042}^{+0.057}$\\
 810.2 & 0.34 & 279.5 &   3 & 0.4 &0.240& 1.043 & $ 0.037_{-0.024}^{+0.030}$\\
 820.1 & 0.36 & 315.2 &   2 & 0.3 &0.244& 1.035 & $ 0.021_{-0.016}^{+0.033}$\\
 840.2 & 0.35 & 677.5 &   8 & 0.8 &0.247& 1.025 & $ 0.042_{-0.016}^{+0.023}$\\
 880.0 & 0.41 & 376.0 &   7 & 0.5 &0.222& 1.001 & $ 0.078_{-0.031}^{+0.045}$\\
 919.9 & 0.44 & 478.6 &   8 & 0.3 &0.256& 0.916 & $ 0.069_{-0.025}^{+0.035}$\\
 939.9 & 0.43 & 469.0 &  22 & 0.7 &0.248& 0.856 & $ 0.214_{-0.047}^{+0.058}$\\
 949.7 & 0.32 & 261.7 &  20 & 0.3 &0.261& 0.855 & $ 0.338_{-0.076}^{+0.095}$\\
 957.7 & 0.32 & 233.9 &  13 & 0.2 &0.263& 0.858 & $ 0.242_{-0.067}^{+0.089}$\\
 969.7 & 0.34 & 251.5 &  29 & 0.5 &0.250& 0.865 & $ 0.524_{-0.099}^{+0.119}$\\
\hline
\end{tabular}
\end{table}
\section{Fitting of the cross section}
\begin{figure}[t]
\centering
\includegraphics[width=0.9\textwidth]{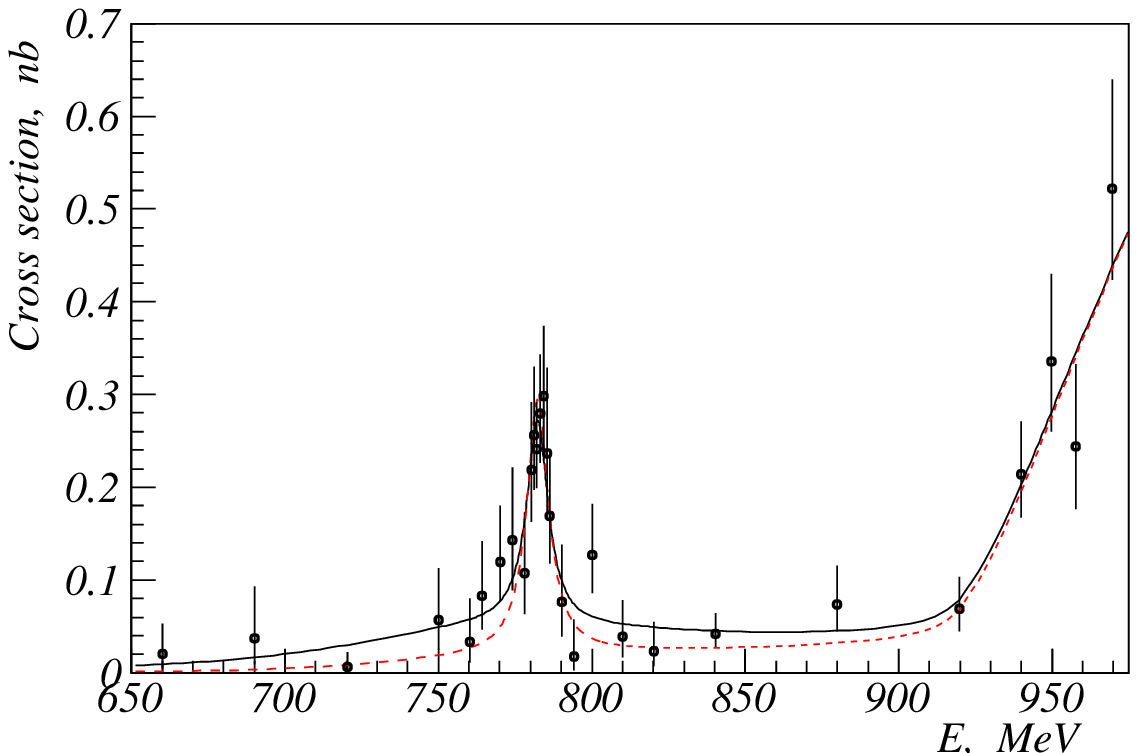}
\caption{\small
The cross section of the $e^+e^-\to\pi^0\pi^0\gamma$ process.
Points with error bars are experimental data.
Solid line is a result of fitting in the model 3 of
Table~\ref{t4}. The dashed line corresponds to the fit with 
$B(\rho\to\sigma\gamma)=0$.
}
\label{f6}
\end{figure}

The fitting procedure maximizes the logarithmic likelihood function
$$L=\sum_i \ln{P_i(N_i^{exp}, N_i^{th})},$$ where $P_i$ is
a Poisson probability to detect observed number of events
$N_i^{exp}$ in the $i$-th energy bin with a theoretical expectation
of $N_i^{th}$. The theoretical expectations
were calculated as
$$N_i^{th}=\varepsilon_i L_i \sigma(E_i) (1+\delta(E_i))+N_i^{bkg},$$
where $N_i^{bkg}$ is a calculated number of background events,
$\varepsilon_i$ is a
detection efficiency, $L_i$ is the integrated luminosity,
$\sigma(E)$ is $e^+e^-\to\pi^0\pi^0\gamma$ cross section depending
on a set of approximation parameters, $\delta$ is
a radiative correction. The radiative correction, which is a functional of
the cross-section energy dependence $\sigma(E)$~\cite{radc}, was determined 
within the fitting procedure.
The values of radiative correction evaluated for each experimental energy
point are listed in
Table~\ref{t1}. The model error of the $(1+\delta)$ value does
not exceed 3\%. The values of the experimental cross section calculated
as
$$\sigma_i^{exp}=\frac{N_i^{exp}-N_i^{bkg}}{\varepsilon_i L_i (1+\delta(E_i))}$$ 
are shown in Fig.\ref{f6} and listed in Table~\ref{t1}.
The systematic error of the cross section is determined by
the the errors of the detector efficiency, integrated luminosity, and
radiative correction. It was estimated to be 7\%.

To calculate the cross section $\sigma(E)$ the amplitude of
the $e^+e^-\to\pi^0\pi^0\gamma$ process was parametrized as
\begin{equation}
A_{\pi\pi\gamma}=A_{\rho\omega\pi}
(BW_\rho+\alpha_1 BW_{\rho^\prime}
+\alpha_2 BW_{\rho^{\prime\prime}})+
 \beta A_{\rho\sigma\gamma}BW_\rho +
\gamma A_{\omega}BW_\omega, \label{ampl}
\end{equation}
The first term in Eq.(\ref{ampl}) is the amplitude
of the $e^+e^-\to \rho,\rho^\prime,\rho^{\prime\prime} \to \omega\pi^0$,
where $\rho^\prime$ and $\rho^{\prime\prime}$
are excitations of the $\rho(770)$.
Second and third terms are
$e^+e^-\to\rho(770)\to\sigma\gamma\to\pi^0\pi^0\gamma$ and
$e^+e^-\to\omega\to\pi^0\pi^0\gamma$ amplitudes.
Each amplitude is written in a factorized form. The functions
$A_{\rho\omega\pi^0}$, $A_{\rho\sigma\gamma}$, $A_{\omega}$
depending on the momenta of final particles
describe the dynamics of vector mesons decays.
The functions $BW_i$
describe the Breit-Wigner resonance shapes:
$$ BW_i=\frac{m_i^2}{m_i^2-E^2-iE\Gamma_i(E)},\;\; i=\rho, \rho^\prime, \rho^{\prime\prime}, \omega.$$  
Here $m_i$ and $\Gamma_i(E)$ are resonance mass and energy dependent width.
The cross section is calculated from Eq.(\ref{ampl}) by integration over the 
phase space of final particles: 
$\sigma(E)=\int |A_{\pi\pi\gamma}|^2 d \Pi$.
At the energy above $\omega\pi$ threshold the Breit-Wigner functions of 
$\rho$ mesons are modified 
$BW_{\rho_i} \to BW_{\rho_i} C_{\rho_i\omega\pi}$,
where $C_{\rho_i\omega\pi}$ are Blatt-Weisskopf factors, restricting
fast growth of the $\Gamma_{\rho_i\omega\pi}$ partial widths~\cite{Clegg}:   
\begin{equation}
C_{\rho\omega\pi}=\frac{1}{\sqrt{1+(Rq_\omega(E))^2}},\:
C_{\rho_i\omega\pi}=
\sqrt{\frac{1+(Rq_\omega(m_{\rho_i}))^2}
{1+(Rq_\omega(E))^2}},\; \rho_i=\rho^\prime,\rho^{\prime\prime}.
\label{Compi}
\end{equation}
Here $q_\omega$ is the $\omega$ meson momentum in $\rho_i\to\omega\pi$ decay.  
The range parameter $R$ is supposed to be the same for
$\rho, \rho^\prime, \rho^{\prime\prime}$ mesons.
The main decay modes of $\rho$ mesons were taken into account for
calculation of the energy dependence of the resonance widths. For instance,
in the case of $\rho(770)$ we use the following expression:
\begin{equation}
\Gamma_{\rho}(E)=\Gamma_{\rho}(m_\rho)
\biggl(\frac{m_\rho}{E}\biggr)^2      
\biggl(\frac{q_\pi(E)}{q_\pi(m_\rho)}\biggr)^3 C_{\rho\pi\pi}^2+ 
\frac{g_{\rho\omega\pi}^2}{12\pi}\, q_\omega^3(E)\, C_{\rho\omega\pi}^2.
\label{grho}
\end{equation}
Here $q_\pi$ is a pion momentum in the $\rho\to 2\pi$ decay.  
The Blatt-Weisskopf factor $C_{\rho\pi\pi}$ is expressed by the formula similar to Eq.(\ref{Compi}).  

\begin{figure}[t]
\centering
\includegraphics[width=0.9\textwidth]{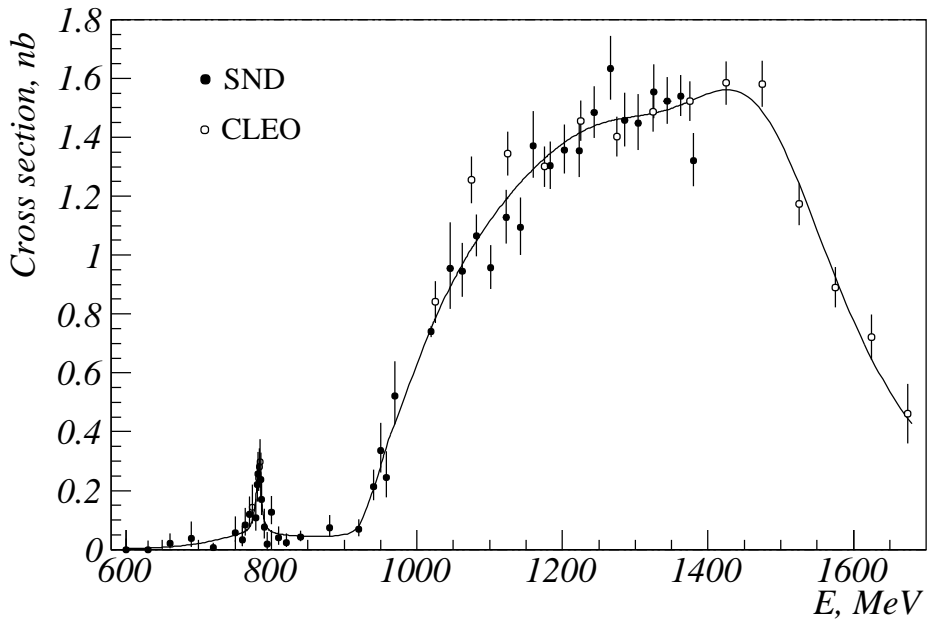}
\caption{\small The cross section of the $e^+e^-\to\pi^0\pi^0\gamma$ reaction.
Points with error bars are experimental data. Curve is a fit result
in the model 3 of Table~\ref{t4}.}
\label{f7}
\end{figure}

The amplitude of the $e^+e^-\to\omega\pi$ process is described by formulas
from Ref.~\cite{ompn,phiompi} and depends on 8 parameters.
The data from energy region below 1~GeV are insufficient to determine them.
Therefore we used additional measurements of the $e^+e^-\to\omega\pi^0$ cross 
section in the 1--1.4~GeV energy range by SND~\cite{phiompi} and 
of the $\omega\pi^0$ spectral function in
$\tau\to\omega\pi\nu_\tau$ decay by CLEO~\cite{CLEOt}.
The spectral function can be converted to corresponding production cross
section in $e^+e^-$ collisions using CVC
hypothesis~\cite{CVC}. The all-data fit on $e^+e^-\to\pi^0\pi^0\gamma$ 
cross section is shown in Fig.\ref{f7}.   
For $E>1\,\mathrm{GeV}$ experimental data are well described
by three models with parameters listed in Table~\ref{t2}.
\begin{table}
\centering
\caption{\small Parameters of the models describing 
$e^+e^-\to\omega\pi\to\pi^0\pi^0\gamma$ cross section for 
$E>1\,\mathrm{GeV}$.\vspace{1mm}}
\label{t2}
\begin{tabular}{|l|c|c|c|c|c|c|c|}
\hline
 & $g_{\rho\omega\pi}$ & $m_{\rho^\prime}$ & $\Gamma_{\rho^\prime}$ & $\alpha_1$ 
 & $m_{\rho^{\prime\prime}}$ & $\Gamma_{\rho^{\prime\prime}}$ & $\alpha_2$ \\
\hline
 1 & 14.3-15.8 & ---  & --- & ---          & 1630-1710 & 630-1000 & -(0.19-0.24) \\
 2 & 14.1-15.7 & 1400 & 500 & -(0.04-0.06) & 1580-1620 & 420-580  & -(0.14-0.18) \\
 3 & 15.4-16.6 & 1400 & 500 & -(0.39-0.42) & 1560-1640 & 380-780  &   0.24-0.30\\
\hline
\end{tabular}
\end{table}
The ranges of parameter values correspond to the variation of the
parameter $R$ from 0 to 2~GeV$^{-1}$. The statistical errors
are not shown because they are significantly smaller than
model biases. For models with two excited $\rho$ states the
$\rho^\prime$ mass and width were fixed to
1400~MeV and 500~MeV. These values are close to
$\rho^\prime$ parameters from $\pi^\pm \pi^0$ spectral function
 data~\cite{ALEPH,CLEO2pi}.  

Below $\omega\pi^0$ threshold the amplitude of the $e^+e^-\to\omega\pi^0$
process drops rapidly and the product
$|BW_\rho|^2\int |A_{\rho\omega\pi}|^2 d\Pi$
in contrast with the corresponding product for $\rho\to \sigma\gamma$ transition
does not demonstrate resonance behavior.
This allows to separate contributions of the two $\rho$ decay mechanisms by
measurement of energy dependence of the $e^+e^-\to\pi^0\pi^0\gamma$ cross
section.
The $\rho\to\sigma\gamma$ decay amplitude was described by
the $\chi PT$ and $L\sigma M$ models from Ref.~\cite{Bramon2}.
The three sets of $\sigma$ parameters
\cite{E791,CLEO,Bramon2} used in the $L\sigma M$ model are
listed in Table~\ref{t4}.
The $\chi PT$ model corresponds to $m_\sigma \to \infty$.
The $\beta$ parameter represents the difference between
observed value of the $\rho\to\sigma\gamma$ decay amplitude and
theoretical prediction.

For $\omega\to\pi^0\pi^0\gamma$ decay the variation of the final state phase
in the CMS energy interval of the $\omega$ meson
is small, so we cannot
separate different decay mechanisms studying the cross
section energy dependence.
Thus the amplitude of the $\omega\to\pi^0\pi^0\gamma$ decay
was written according to VMD model~\cite{ompn}. The $\rho$-$\omega$ mixing
was taken into account following Ref.~\cite{Bramon2}. Possible
contributions of other mechanisms would result in a deviation of the
complex parameter $\gamma$ from 1.

 Full description of the energy dependence of the cross
section below 1~GeV requires extra four parameters e.g.
absolute values and phases of $\beta$ and $\gamma$:
$|\beta|$, $\phi_\beta$, $|\gamma|$, $\phi_\gamma$.
But we prefer two other sets of parameters:
$B(\rho\to\sigma\gamma)$, $\phi_\beta$,
$B(\omega\to\pi^0\pi^0\gamma)$, $\phi_\gamma$ or
$B(\rho\to\pi^0\pi^0\gamma)$, $\phi_\beta$, $B(\omega\to\pi^0\pi^0\gamma)$,
$\phi_\gamma$. The branching ratios are related to $\beta$ and
$\gamma$ as
$$
B(\rho\to\pi^0\pi^0\gamma)=\frac{m_\rho^2}{\Gamma_\rho^2}
\frac{1}{\sigma_\rho} \int |A_{\rho\omega\pi}(m_\rho)+
\beta A_{\rho\sigma\gamma}(m_\rho)|^2 d\Pi, $$
$$B(\rho\to\sigma\gamma)=\frac{m_\rho^2}{\Gamma_\rho^2}
\frac{1}{\sigma_\rho} \int |
\beta A_{\rho\sigma\gamma}(m_\rho)|^2 d\Pi,$$
$$B(\omega\to\pi^0\pi^0\gamma)=\frac{m_\omega^2}{\Gamma_\omega^2}
\frac{1}{\sigma_\omega}\int |
\gamma A_{\omega}(m_\rho)|^2 d\Pi,$$
where $\sigma_V=12\pi B(V\to e^+e^-)/M_V^2$ is a total 
vector meson production cross section in $e^+e^-$ collisions.

Characteristic feature of the process under study is a large
interference between the contributions of $\rho$ and
$\omega$ decays. For instance the cross section of
$e^+e^-\to\omega\to\pi^0\pi^0\gamma$ process at $E=m_\omega$ evaluated using
the table value of
$B(\omega\to\pi^0\pi^0\gamma)=(7.2\pm 2.5)\cdot10^{-5}$
is equal to 0.12 nb. The interference with $\rho$ meson increases
this value up to approximately 0.25 nb (Fig.\ref{f6}).
The experimental data on the energy dependence of the cross section
are insufficient for determination of unambiguous solution for interference phases
$\phi_\beta$, $\phi_\gamma$. There are four solutions listed
in Table~\ref{t3}.
\renewcommand{\arraystretch}{1.3}
\begin{table}
\centering
\caption{\small
The branching ratios of $\rho$ and $\omega$ decays
($B\times 10^{5}$) and $P(\chi^2)$ values obtained as a result
of cross section fitting with different values of
$\phi_\beta$, $\phi_\gamma$. The $\omega\pi$ amplitude
was described by the
Model 3 from Table~\ref{t2}. Model 1 from Table~\ref{t4}
was used for description of $\rho\to\sigma\gamma$ 
amplitude. Only statistical errors of parameters are shown.
\vspace{1mm}}
\label{t3}
\begin{tabular}{|c|c|c|c|c|c|c|c|}
\hline
\multicolumn{2}{|c|}{Model} & $B(\omega\to\pi^0\pi^0\gamma)$ & $B(\rho\to\pi^0\pi^0\gamma)$ & $B(\rho\to\sigma\gamma)$ & $P(\chi^2)$\\
\hline
1&  $\phi_\beta\approx 0$, $\phi_\gamma\approx 0$ & $6.3_{-1.3}^{+1.4}$ & $4.1_{-0.9}^{+1.0}$ & $1.9_{-0.8}^{+0.9}$ & 35\% \\
2&  $\phi_\beta\approx \pi$, $\phi_\gamma\approx 0$ & $12.3_{-1.6}^{+2.3}$ & $3.8_{-0.8}^{+0.9}$ & $4.4\pm1.0$ & 30\%\\
3&  $\phi_\beta\approx 0$, $\phi_\gamma\approx \pi$ & $25.5_{-2.3}^{+2.4}$ & $5.1_{-0.9}^{+1.0}$ & $1.9\pm1.0$ & 6\%\\
4&  $\phi_\beta\approx \pi$, $\phi_\gamma\approx \pi$ & $15.8\pm 2.3$ & $4.7_{-0.8}^{+0.9}$ & $5.6_{-1.0}^{+1.1}$ & 8\% \\
\hline
\end{tabular}
\end{table}
The third and fourth ones correspond to a large
destructive contribution into $\omega$ decay from mechanisms other than $\rho^0\pi^0$.
The $B(\omega\to\pi^0\pi^0\gamma)$ values obtained in this
case disagree with existing experimental value
$B(\omega\to\pi^0\pi^0\gamma)=(7.2\pm 2.5)\cdot10^{-5}$.
The solution with $\phi_\beta\approx \pi$, $\phi_\gamma\approx 0$
can be ruled out for two reasons: $B(\omega\to\pi^0\pi^0\gamma)$ exceeds 
the table value by 1.7 standard deviations and consistency of the calculated
spectrum of the recoil photon with the 
experimental one is poor. The analysis of the photon spectrum is described 
in the next section.

For the only survivor solution with both phases close to zero, 
the model dependence
of the fit parameters was studied. Three models of excited $\rho$ states
(Table~\ref{t2}) and four sets of $\sigma$ parameters (Table~\ref{t4})
were tested. The $\phi_\beta$ was found ranging within
$(20\div 80)^\circ \pm 80^\circ$. These values are in agreement with theoretically
expected zero value~\cite{Bramon2}. Therefore the final fitting was
performed with $\phi_\beta=0$. The phase $\phi_\gamma$ was considered as
a floating parameter to take into account its possible shift due
to the contribution of mechanisms other than $\omega\to\rho\pi^0$.
The fitted $\phi_\gamma=-(2\div 20)^\circ\pm 20^\circ$ is
consistent with zero.

The probabilities of the $\rho$ and $\omega$ decays into $\pi^0\pi^0\gamma$
obtained with different parameters of $\sigma$ meson are listed
in Table~\ref{t4}.
\renewcommand{\arraystretch}{1.3} 
\begin{table}
\centering
\caption{\small
The probabilities of the $\rho$ and $\omega$ decays into $\pi^0\pi^0\gamma$
($B\times 10^{5}$)
obtained with different parameters of $\sigma$ meson.
The spreads in the parameter values correspond to the models listed
in Table~\ref{t2}. Theoretical values of branching ratios are taken from
Ref.~\cite{Bramon2}.
The final results with statistical and systematic errors are given in the 
bottom line of the table. \vspace{1mm}
} 
\label{t4}
\begin{tabular}{|c|c|c|c|c|c|c|c|}
\hline
 Model& $m_\sigma$ & $\Gamma_\sigma$ &
 $B(\omega\to\pi^0\pi^0\gamma)$ &
 \multicolumn{2}{c|}{$B(\rho\to\pi^0\pi^0\gamma)$} &
 \multicolumn{2}{c|}{$B(\rho\to\sigma\gamma)$} \\
 \cline{5-8}
&  MeV &  MeV & & exp.  & th. & exp.  & th. \\
\hline
  L$\sigma$M & 478 & 324  & 6.2-6.5 & 4.0-4.1  & 3.8 & 1.5-1.9 & 1.5 \\
  L$\sigma$M & 555 & 540  & 6.3-6.6 & 4.2-4.3  & 2.8 & 1.9-2.3 & 0.8 \\
  L$\sigma$M & 478 & 263  & 6.2-6.4 & 4.2-4.3  & 4.7 & 1.7-2.1 & 2.1 \\
  $\chi$PT   & --- & ---  & 6.5-6.9 & 3.9-4.0  & 2.9 & 1.8-2.2 & 1.0 \\
\hline
\multicolumn{3}{|c|}{ }& 
\multicolumn{1}{c|}{$6.6_{-1.3}^{+1.4}\pm0.6$}&
\multicolumn{2}{c|}{$4.1_{-0.9}^{+1.0}\pm0.3$}&
\multicolumn{2}{c|}{$1.9_{-0.8}^{+0.9}\pm0.4$}
\\
\hline
\end{tabular}
\end{table}
The spreads in parameter values correspond to different
models describing $e^+e^-\to\omega\pi^0$ cross section
above 1~GeV. All models reproduce the experimental data well.
Therefore parameter midrange was taken as a final result.
Its spread was regarded as the model error. The branching
ratios obtained this way with statistical and systematic
errors are listed in the last
row of Table~\ref{t4}.
The systematic error includes the model error,
uncertainties in the detection efficiency and integrated luminosity.
The variation of the background level within its systematic
error practically does not change the $B(\omega\to\pi^0\pi^0\gamma)$
central value and results in following additional uncertainties of the
$\rho$ meson branching ratios: 7\% for $B(\rho\to\pi^0\pi^0\gamma)$ and 12\% for
$B(\rho\to\sigma\gamma)$. Since these uncertainties affect
statistical significance of the results they were added to statistical errors.
The energy dependence of the cross section in the model with $m_\sigma=478$~MeV
and $\Gamma_\sigma=324$~MeV is shown in Fig. \ref{f6} together with the
curve corresponding to $B(\rho\to\sigma\gamma)=0$.
The $P(\chi^2)$ value for the latter model is equal to 0.5\%.
\section{The energy and angular spectra}
From Table~\ref{t1} it is seen that selected events are mainly
concentrated in two energy regions:
180 events near $\omega$ peak and 92 events in the range 920--970~MeV
above the
$e^+e^-\to\omega\pi^0$ reaction threshold. The angular and energy
distributions in the  latter region agree with $\omega\pi^0$
mechanism.
\begin{figure}[t]
\begin{minipage}[t]{0.47\textwidth}
\includegraphics[width=0.98\textwidth]{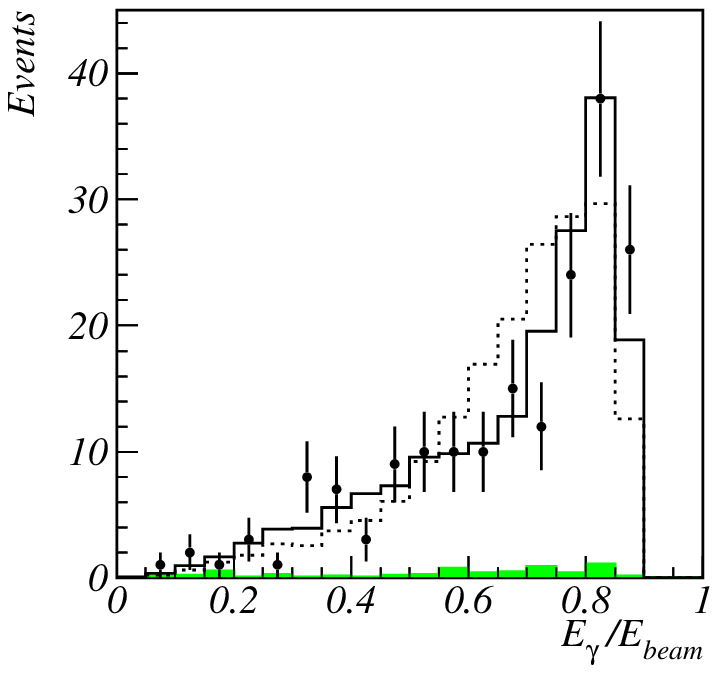}
\caption{\small 
The recoil photon spectrum for the experimental events of
the reaction (\ref{ppg}) in the energy range $760<E<800$~MeV
(points with error bars) and the results of simulation
in the model 3 from Table~\ref{t4} (solid line) and
in the model 2 from Table~\ref{t3} (dashed line).
}
\label{f3}
\end{minipage}
\hfill
\begin{minipage}[t]{0.47\textwidth}
\includegraphics[width=0.98\textwidth]{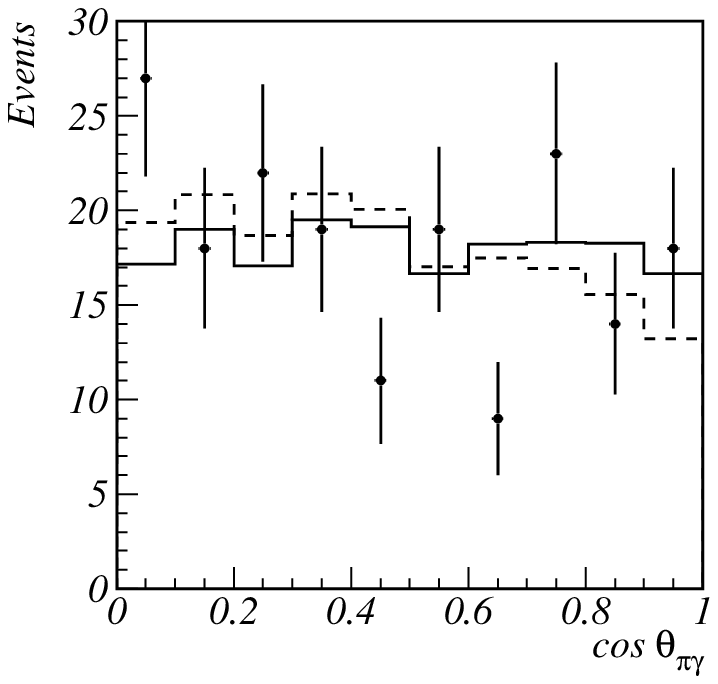}
\caption{\small
Distribution of cosine angle between photon
and $\pi^0$ meson in the $\pi^0\pi^0$ rest frame.
Point with error bars are experimental data,
Solid line is a simulation with $\omega$ decay into pure
$\rho^0\pi^0$ state. Dashed line is a simulation with $\omega$
decaying into a mixture of $\rho^0\pi^0$ and $\sigma\gamma$ states.
}
\label{f4}
\end{minipage}
\end{figure}
Recoil photon spectrum for events from 760--800~MeV energy range is shown in
Fig.\ref{f3}. Although this energy region is
dominated by $\omega$ peak the contributions of both $\omega$ and $\rho$
decays must be taken into account to obtain the theoretical spectrum.
The spectrum calculated in model 3 from Table~\ref{t4} (Fig.\ref{f3})
is in a good agreement with the experimental one. In this model
it is supposed that $\omega\to\pi^0\pi^0\gamma$ decay proceeds 
through $\rho^0\pi^0$ intermediate state.
Another way is to describe the $\omega$ decay by a sum of contributions
of $\omega\to\rho^0\pi^0$ and $\omega\to\sigma\gamma$ mechanisms.
To do this we fix $B(\omega\to\rho^0\pi^0\to\pi^0\pi^0\gamma)$ at
$2.5\cdot 10^{-5}$ and fit the
$\omega\to\sigma\gamma$ decay contribution to a value yielding
observed $\omega\to\pi^0\pi^0\gamma$ branching ratio.
In case of constructive interference this leads to
$B(\omega\to\sigma\gamma\to\pi^0\pi^0\gamma)=1.3\cdot 10^{-5}$
and the photon spectrum close to that expected for $\rho^0\pi^0$ mechanism.
As was shown in the previous section, assumption of destructive interference
results in $B(\omega\to\pi^0\pi^0\gamma)$
inconsistent with the PDG table value.

The second theoretical spectrum in Fig.\ref{f3} corresponds to
the model 2 from Table~\ref{t3} with destructive interference
of $\omega\pi$ and $\sigma\gamma$ amplitudes in  $\rho$ decay.
For this model the consistency  between
theoretical and experimental spectra calculated using Kolmogorov
test \cite{hbook} is about 1\%,
which was one of the reasons to discard this model.

Additional information about mechanism of $\omega$ decay
can be obtained from the analysis of angular distributions.
One of such distribution is shown in Fig.\ref{f4}.
The same figure displays the theoretical distributions
obtained under assumptions that $\omega$ decay proceeds
through either pure $\rho^0\pi^0$ intermediate state
or a mixture of $\rho^0\pi^0$ and $\sigma\gamma$.
One can see that our limited statistics does not allow to
distinguish these two models.
\section{Summary}
The branching ratios measured in this work, 
\begin{equation}
B(\omega\to\pi^0\pi^0\gamma)=(6.6_{-1.3}^{+1.4}\pm0.6)\cdot10^{-5},
\label{oppg}
\end{equation}
\begin{equation}
B(\rho\to\pi^0\pi^0\gamma)=(4.1^{+1.0}_{-0.9}\pm0.3)\cdot10^{-5}.
\label{rppg}
\end{equation}
are in a good agreement with our preliminary results~\cite{rhoppg} 
and GAMS measurement  
$B(\rho\to\pi^0\pi^0\gamma)=(7.2\pm2.5)\cdot10^{-5}$~\cite{GAMS}, 
but have higher accuracy.

The probability of $\rho\to\pi^0\pi^0\gamma$ decay significantly exceeds
VMD model prediction $(1.3-1.5)\cdot10^{-5}$. This excess can
be explained by the contribution of the decay via scalar state
$\rho\to\sigma\gamma$. Two mechanisms, $\rho\to\sigma\gamma$ 
and $\rho\to\omega\pi$, can be separated using difference in energy dependence of
their amplitudes.
Our result on $\rho\to \sigma\gamma$ decay 
\begin{equation}
B(\rho\to \sigma\gamma\to\pi^0\pi^0\gamma)=(1.9_{-0.8}^{+0.9}\pm0.4)\cdot10^{-5}
\end{equation}
differs from zero by 2.4 standard deviations and is consistent
with the predictions of chiral models~\cite{Bramon2,Oset}.
The magnitude of $B(\rho\to \sigma\gamma)$ is sensitive to
$\sigma$ parameters. As can be seen from Table~\ref{t4},
the models with $\Gamma_\sigma\sim300$~MeV give the most
consistent description of the experimental data.

The value of the branching ratio
$B(\omega\to\pi^0\pi^0\gamma)=(6.7\pm1.2)\cdot10^{-5}$,
obtained by averaging of our measurement with the GAMS result
exceeds theoretical predictions,
$(4.6\pm1.1)\cdot10^{-5}$~\cite{Bramon2,Singer2} and
$(4.7\pm0.9)\cdot10^{-5}$~\cite{Oset},
by 1.3 standard deviations.
It is necessary to make some remarks about these predictions.
The result of Ref.~\cite{Singer2} is based on table value of
$\Gamma(\rho\to\pi^0\gamma)=102\pm26$ keV~\cite{PDG2000}. It must be
corrected taking into account newer measurement
$\Gamma(\rho\to\pi^0\gamma)=76\pm22$ keV~\cite{p0g}, which is close
to the value for charged $\rho$, $\Gamma(\rho^\pm\to\pi^\pm\gamma)=68\pm8$ keV~\cite{PDG2000}.
This decreases the predicted $B(\omega\to\pi^0\pi^0\gamma)$
and worsens agreement with the experiment.
In the Refs.~\cite{Bramon2} and \cite{Oset} the values of $g_{\rho\omega\pi}$
equal to 15~GeV$^{-1}$ and  15.9~GeV$^{-1}$ were used to
calculate
$B(\omega\to\pi^0\pi^0\gamma)\propto g_{\rho\omega\pi}^2
g_{\rho\pi\gamma}\propto g_{\rho\omega\pi}^4/g_{\rho}^2$.
On the other hand, the use of these $g_{\rho\omega\pi}$ values for VMD
calculation of $\Gamma(\omega\to 3\pi)$ and $\Gamma(\omega\to \pi^0\gamma)$
leads to too large values conflicting with experimental data.
For example, the $g_{\rho\omega\pi}$ obtained
from $\omega\to 3\pi$ width assuming intermediate $\rho\pi$
state is equal to $(14.3\pm0.2)$~GeV$^{-1}$~\cite{ompn}.
Therefore, our opinion is that $4.6\cdot 10^{-5}$
is the maximum branching ratio acceptable within VMD model
and the additional theoretical study is required to
explain the large value of $B(\omega\to\pi^0\pi^0\gamma)$.
\section{Acknowledgments}
This work is supported by ``Russian Fund for Basic Researches''
grants No. 01-02-16934-a and 00-15-96802
and STP ``Integration'' Fund, grant No. A0100.  


\begin {thebibliography}{20}
\bibitem{SND}
M.N.~Achasov et al., 
Nucl.~Instr.~Meth. A 449 (2000) 125.
\bibitem{rhoppg}M.N.~Achasov et al., JETP Letters, 71 (2000) 355.
\bibitem{GAMS}D.~Alde et al., Phys.~Lett. B 340 (1994) 122.
\bibitem{Singer1}P.~Singer, Phys.~Rev. 128 (1962) 2789.
\bibitem{Bramon1}A.~Bramon, A.~Grau, G.~Pancheri, Phys.~Lett. B 283 (1992) 416.
\bibitem{Bramon3}A.~Bramon, A.~Grau, G.~Pancheri, Phys.~Lett. B 289 (1992) 97. 
\bibitem{Bramon2}A.~Bramon et al., Phys.~Lett. B 517 (2001) 345.
\bibitem{Marco}E.~Marco et al., Phys.~Lett. B 470 (1999) 20.
\bibitem{Oset}J.~E.~Palomar, S.~Hirenzaki, E.~Oset, e-print hep-ph/0111308.
\bibitem{Gokalp1}  A.~Gokalp, O.~Yilmaz, Phys.~Lett. B 508 (2001) 25.
\bibitem{Singer2}D.~Guetta, P.~Singer, Phys.~Rev.  D 63 (2001) 017502.
\bibitem{Gokalp2}  A.~Gokalp, O.~Yilmaz, Phys.~Lett. B 494 (2000) 69.
\bibitem{xinm}A.V.~Bozhenok, V.N.~Ivanchenko, Z.K.~Silagadze,
Nucl.~Instr.~Meth. A 379 (1996) 507.
\bibitem{ompn}
M.N.~Achasov et al., 
Nucl.~Phys. B 569 (2000) 158.  
\bibitem{radc}
E.A.~Kuraev, V.S.~Fadin, Sov. J. Nucl. Phys. 41 (1985) 466. 
\bibitem{Clegg}A.B.~Clegg, A.~Donanachie, Z.~Phys. C 62 (1992) 455.   
\bibitem{phiompi} M.N.~Achasov et al., Phys.~Lett. B 486 (2000) 29.
\bibitem{CLEOt} K.W.~Edwards et al.,
Phys.~Rev. D 61 (2000) 072003.
\bibitem{CVC}Y.S.~Tsai, Phys. Rev. D 4 (1971) 2821.
\bibitem{ALEPH}R.~Barate et al.,
Z.~Phys. C 76 (1997) 15.
\bibitem{CLEO2pi}S.~Anderson et al.,
Phys. Rev. D 61 (2000) 112002.
\bibitem{E791} E.M.~Aitala et al.,
Phys.~Rev.~Lett. 86 (2001) 770.
\bibitem{CLEO} D.M.~Asner et al.,
Phys. Rev. D 61 (2000) 012002.
\bibitem{hbook}HBOOK Reference Manual, CERN program library Y250, 1998, p.90. 
\bibitem{PDG2000} Review of Particle Physics. 
Eur. Phys. J. C 15 (2000) 1.
\bibitem{p0g} M.N.~Achasov et al., e-print hep-ex/0109035. 
\end {thebibliography}
\end{document}